# Evaluation of a photonic lantern spatial demultiplexer based receiver for optical communication

Vincent Billault,[1,*] Luc Leviandier,[1] Jérome Bourderionnet,[1] Christophe Pierre,[2] Raynald Sanquer,[2] Marc Castaing,[2] and Arnaud Brignon[1]

[1]Thales Research and Technology France, 1 Avenue Augustin Fresnel, 91767 Palaiseau, France
[2] Alphanov, rue François Mitterrand, 33400 Talence, France
* vincent.billault@thalesgroup.com

**In this paper, we present a method based on a modal decomposition to quantify the efficiency of photonic lantern (PL) based free space optical (FSO) communication receivers. We fabricate a seven port PL and we evaluate numerically the free space to fiber coupling efficiency based on a reconstruction of the fields at the PL FSO multimode port. We validate the numerical approach with an experimental characterization of the PL. Then we compare the PL to a commercial multiplane light converter spatial demultiplexer. The PL shows better coupling efficiency for low order spatial modes with orders of magnitude of demultiplexer size reduction. Finally we evaluate the PL receiver with a simulation of received optical wavefront in a FSO communication.**

Coupling free space optical (FSO) wavefronts after propagation through atmosphere in single mode fibers (SMF) is a crucial step for FSO communications [1]. As atmospheric turbulence degrade the spatial quality of the beam through its propagation [2], the optical receiver needs to be carefully designed and adapted to minimize the impact of the degradation on the optical budget link and on the communication signal. Adaptive optics is commonly used to compensate for atmospheric turbulence [3]. A wavefront sensor and a deformable mirror restore the beam quality at the reception over time in order to maintain a good free space to fiber coupling efficiency.

As an alternative, FSO space-division demultiplexers project a multimode FSO wavefront in a set of single mode beams coupled in SMFs with variable amplitude and phase [4-5]. The same communication signal encoded on each optical carrier propagating through the SMFs can thus be summed in the electrical [6-7] or optical [8-9] domain to mitigate the amplitude and phase variations of the wavefront over time. The free space to fiber coupling losses and the fraction of the spatial wavefront coupled in the SMFs are two key parameters determining the overall efficiency of the demultiplexer for communications.

Multi-apertures receivers (with or without independent tip/tilt correction) can be used as spatial demultiplexers [4, 10-11], ensuring a good free space to fiber coupling, however at the detriment of a bulky receiver with a very low fraction of received wavefront coupled in the SMFs. On the other hand, mutli-plane light conversion (MPLC) modules project a FSO wavefront on a mode basis ensuring an almost complete collection of the distorted wavefront, but at the expanse of a higher free space to fiber insertion losses and a complicated assembly in a rather bulky packaging [12-13].

Alternatively, photonics lanterns (PL) are good candidates for FSO communication demultiplexer as they are broadband and compact devices [14-15]. PL consist in bundle of SMF tapered to a multimode fiber section [16]. The PLs free space to fiber insertion losses can be very low and over the past decades PLs have proven their interest for optical communication, increasing the overall performances of the receivers [7,8,17-18]. However not so much attention is given to a more generic method to evaluate the efficiency of PL receivers for FSO communication. In this paper, we use a modal approach to quantify the efficiency of a PL based receiver for FSO communications. The modal coupling efficiency is evaluated numerically and experimentally for a seven port PL. The PL is then compared numerically and experimentally to a MPLC spatial demultiplexer with a simulation of received optical wavefront in a FSO communication scenario.

A PL is composed of different sections (Fig. 1): a bundle of SMFs (packed SMF) placed inside a capillary are tapered into a multimode (MM) section [16]. For FSO communications, the MM section is cleaved to have a free space/MM interface. By stretching the packed SMF, the size of the fiber diameter and fiber core is reduced homothetically. The mode field diameter (MFD) of each SMF increases as the fiber diameter reduces until the different mode start to couple with the adjacent SMF's modes. To quantify the ability of a PL to couple FSO wavefronts to SMFs, we measure the different fields $\psi_i(x, y)$ in the output MM section when the light is injected though the different PL SMF ports. Then, for each of these fields, we evaluate the coupling efficiency matrix $\eta_{(m,n)}^{i}$ corresponding to the power modal decomposition on the Hermite Gauss (HG) basis $\phi_{m,n}(x, y)$ (all the fields considered are normalized).

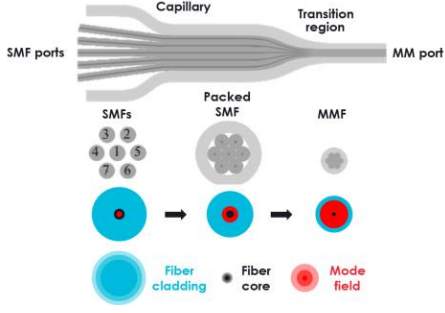

**Fig. 1** Schematic of the PL design.

$$\eta_{m,n}^i = \left| \iint \psi_i(x,y) \cdot \phi_{m,n}^*(x,y) dx dy \right|^2. \quad (1)$$

We chose particularly the HG mode basis as it corresponds to a compact basis of FSO propagation. If the orthogonality of the fields $\psi_i$ is assessed, we derive the PL modal coupling efficiency for a HG mode $\eta_{m,n}$ by summing the coupling efficiency on each field $\psi_i$:

$$\eta_{m,n} = \sum_i \eta_{m,n}^i. \quad (2)$$

$\eta_{m,n}$ is then the overlap integral of the HG mode and its projection on the fields space of the PL, generated by the basis $\{\psi_i(x,y)\}$. This modal coupling efficiency approach can be used to compare FSO spatial demultiplexers both in term of FSO to fiber coupling losses, and collection of the distorted wavefront (number of modes collected).

A seven port PL is fabricated with stretching/ splicing of seven SM1500G80 SMFs. The initial fiber cladding diameter is 80 μm and is reduced to ~25 μm in the MM section. The MM section is cleaved and its total diameter is 71 μm. The PL SMF ports are spliced to SMF28 fibers to inject light in the PL through the SMF ports and to measure the light coupled from the MM FSO port. To evaluate the fields $\psi_i(x,y)$ we measure the near field intensity using an afocal doublet with magnification of 10, and the far field intensity of the MM FSO port of the PL, using a single lens, when injecting light in the different FSO ports (Fig. 2). The fields $\psi_i(x,y)$ are reconstructed based on the near field and far field imaging numerically in amplitude and phase with a Gerchberg-Saxton (GS) algorithm (Fig. 2). The overlap integral between the amplitude measured in the near field $|\psi_{i,\exp}(x,y)|$ and the amplitude of the field reconstructed $|\psi_i(x,y)|$ via the GS algorithm [19] for the 7 SMF ports is 0.97 ±0.004, owing for a good reconstruction of the fields. We compute the matrix of the square modulus of the overlap integral between the fields $\psi_i$ (Fig. 3.a). The maximum of its non-diagonal elements is −18.4 dB. The orthogonality of the fields injected via the SMF ports of the PL is therefore preserved all along the PL.

We then derive the square modulus of the projection of the HG mode basis $\phi_{m,n}$ on the fields $\psi_i$ to evaluate $\eta_{m,n}^i$ (Eq.1) and the modal coupling efficiency $\eta_{m,n}$ (Fig. 3.b). The fundamental HG$_{00}$ mode waist is set at 18 μm to maximize the coupling of this fundamental mode to the PL. The PL shows a good coupling efficiency for low order HG modes $(m + n)$: 0.8 dB losses for HG$_{00}$ mode and ≤ 4 dB losses for 7 HG mode, namely $(m,n) = (0,0), (0,1), (1,0), (0,2), (1,1), (2,0), (1,2)$.

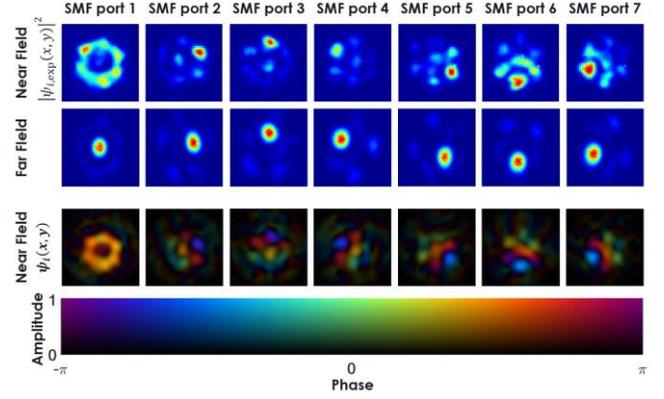

**Fig. 2** Near (top) and far (mid) field intensity imaging of the PL MM section when injecting light on the different SMF ports. Fields $\psi_i(x,y)$ reconstructed with GS algorithm (bot).

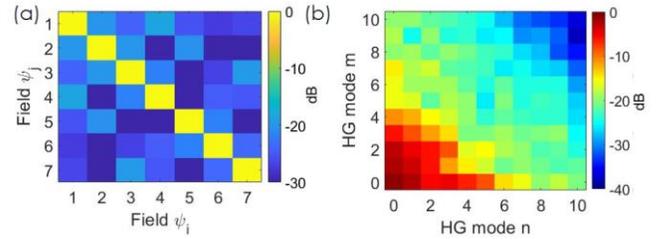

**Fig. 3** (a) Overlap integral matrix between the fields reconstructed $\psi_i$ at the MM FSO port. (b) PL modal coupling efficiency evaluated numerically (Eq. 2).

The losses however increases as the HG mode order increases because the energy in the high-order modes outer lobes is less likely guided by the cores of the different fibers in the PL MM part. In contrast, low-order modes couple to the different cores of the SMF fibers through the propagation in the PL.

We measure experimentally $\eta_{m,n}$ to validate the numerical approach based on the reconstructed fields $\psi_i$. A free space wavefront emulator setup (Fig. 4) generate HG modes on the PL FSO MM port. The wavefront emulator consists of a 1550 nm laser illuminating a spatial light modulator (SLM). The SLM generates spatially modulated wavefronts in both amplitude and phase by superimposing arbitrary profiles onto a diffraction grating and retaining only the 1st diffraction order using a pinhole [20]. The diffracted light is imaged by a camera via an afocal doublet (with a magnification of 0.8), and a second afocal doublet (with a magnification of 0.05) images the SLM onto the FSO MM port of the PL. The optical power coupled into each PL SMF port is measured simultaneously with a powermeter to evaluate the matrix $\eta_{m,n}^i$. The size and position of the fundamental HG$_{00}$ mode have been optimized to maximize the output power in the PL SMF ports. The PL experimental modal coupling efficiency $\eta_{m,n}$ is represented in Fig. 5. The simulation agrees well with the experimental results. High-order HG$_{mn}$ modes (with m+n>2) have lower losses than low-order modes (with m+n<2). The PL shows a very good coupling efficiency (>45%) for the first six HG modes.

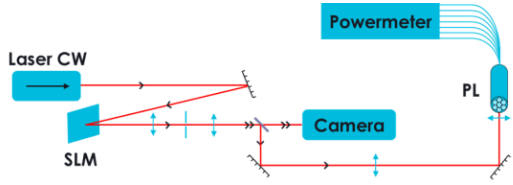

**Fig. 4** Free space photonic lantern characterization setup. SLM: spatial light modulator. PL: photonic lantern.

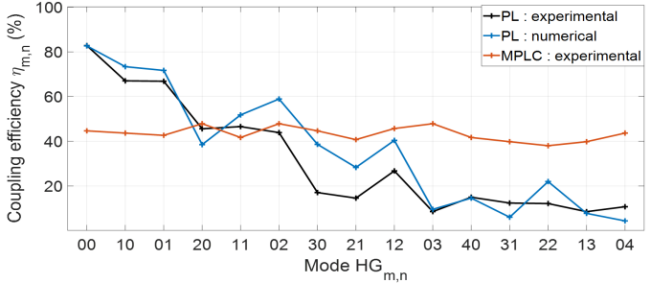

**Fig. 5** PL Modal coupling efficiency $\eta_{m,n}$: experimental (black) and based on the numerical power projection of the HG mode on the fields $\psi_i$ (blue). Experimental modal coupling efficiency of a commercial MPLC.

We evaluate the modal insertion losses of a commercial MPLC (Fig. 5) designed for 15 HG modes with the same setup. The MPLC has a uniform modal coupling efficiency, but the PL shows a lower losses for low order HG modes with only 7 SMF ports.

We verify experimentally the PL linearity to ensure that the PL modal decomposition approach can be used with a distribution of HG modes with variable amplitude and phase. We generate a superposition of two HG modes ($HG_{01}$ and $HG_{10}$, that have similar modal coupling efficiency) with the FSO wavefront emulator and we change the relative amplitude and phase between the two modes. We define $\sqrt{\rho}$ the amplitude of the mode $HG_{01}$ (the amplitude of the mode $HG_{10}$ is $\sqrt{1-\rho}$) and $\delta\varphi$ the phase offset between the two modes. The total field displayed on the SLM screen and injected in the FSO MM port of the PL $\phi(x,y)$ writes:

$$\phi(x,y) = \sqrt{\rho}\,\phi_{01}(x,y) + e^{i\delta\varphi}\sqrt{1-\rho}\,\phi_{10}(x,y). \quad (3)$$

We define $\eta^i$ as the ratio of the power in the $i^{th}$ SMF port versus the FSO power injected in the PL and $\eta^{tot}$ as the sum of the power in the SMF fibers versus the FSO power injected in the PL (Fig. 6):

$$\eta^i = \left|\iint \psi_i(x,y) \cdot \phi^*(x,y) dx dy\right|^2. \quad (4)$$

The optical power coupled in the SMF ports output varies with $\rho$ and $\delta\varphi$ due to the interference of the projections of the mode $HG_{01}$ and $HG_{10}$ on the individual SMF ports. The SMF port 1 output has up to 20 dB of coupling loss variation when $\delta\varphi$ goes from 0 to $\pi$. However, the total PL coupling loss has less than 1 dB variation when we change $\delta\varphi$ for every $\rho$. Therefore, apart from the inherent modal PL losses $\eta_{m,n}$ the energy is conserved between the MM port and the SMF ports when we inject a superposition of modes with different amplitude and phase.

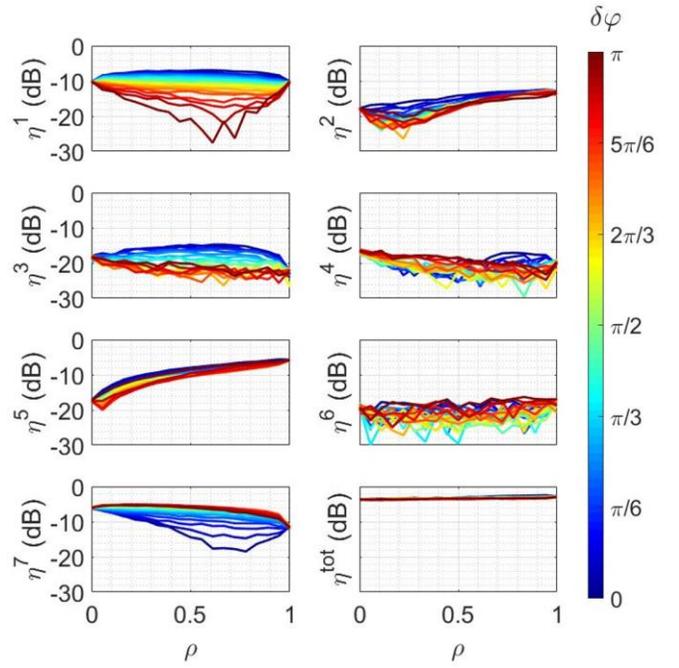

**Fig. 6** Experimental results showing the PL coupling loss $\eta^i$ for a superposition of mode $HG_{01}$ and $HG_{10}$ in the FSO MM input port as a function of the relative amplitude $\rho$ and the phase offset $\delta\varphi$.

The modal coupling efficiency $\eta_{m,n}$ approach can also be used to estimate the coupling losses of complex fields on the PL.

Finally, we evaluate the PL receiver for a GEO satellite to an optical ground station (OGS) communication with $\eta_{m,n}$ numerically estimated and experimentally measured. We use a numerical model of optical wave propagation in a turbulent atmosphere to generate times series of received wavefront. This model use a beam propagation method with a series of FSO propagation, evaluated in Fourier space, and phase screens accounting for refraction indices fluctuations. More details about the numerical model hypothesis can be found in [21]. The fields in the receiver plane of the time series issued from the numerical model are projected onto the 15 first Hermite-Gauss (HG) modes (i.e. the HG modes in Fig. 5). Fig. 7 shows an example of the relative power distribution of the received wavefront for the 15 first HG modes for the 100 first frames generated by the numerical model.

We estimate the PL receiver losses over time for the GEO to OGS FSO link by multiplying the received wavefront power distribution with $\eta_{m,n}$ (Fig. 8 in blue). The PL receiver losses are also measured experimentally (Fig. 8 in black) by displaying the received wavefronts from the time series on the SLM. The PL receiver losses is then the ratio of the sum of the power in the SMF fibers versus the FSO power injected in the PL FSO MM port. The simulation agrees well with the experimental results, which also validate the modal coupling efficiency approach to qualify a PL. The PL receiver shows a $-3.3$ dB coupling losses in average with limited fadings (4.2 dB of coupling efficiency variation).

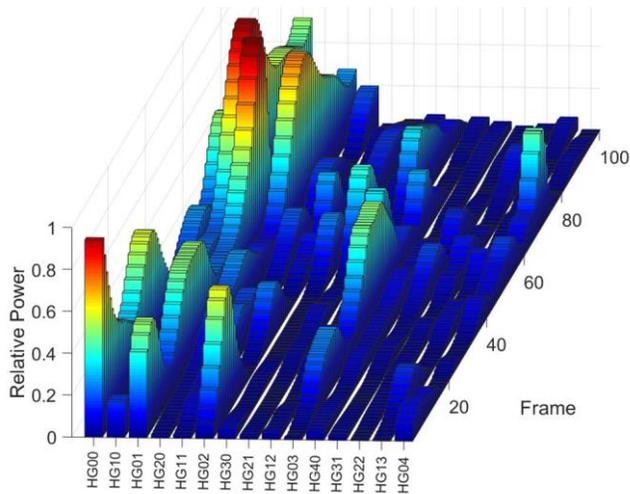

**Fig. 7** Relative power distribution of the first 15 HG modes versus time for the GEO to OGS FSO link.

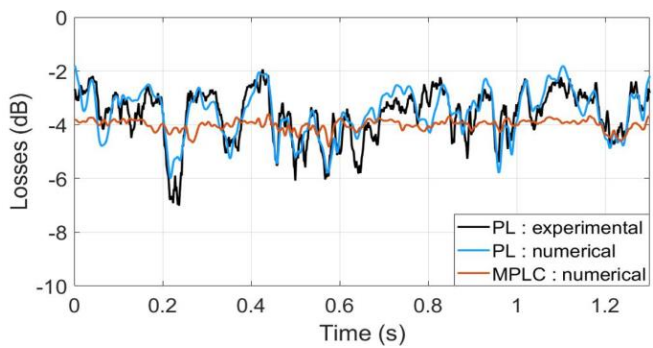

**Fig. 8** Evolution the spatial demultiplexer losses versus time for an incoherent combination of the SMF fibers output.

Compared to the response of a commercial spatial demultiplexer (estimated numerically based on the modal coupling efficiency $\eta_{m,n}$, Fig. 8 in red), the PL receiver has a 1 dB lower coupling losses in average but higher fadings. However, the PL fabricated has only 7 SMF ports compared to the 15 HG modes commercial spatial demultiplexer. As the PL shows lower losses for low order HG modes, we expect PL receivers with a higher number of SMF port to have lower average loss together with lower output power variation.

We have proposed a method based on a modal decomposition to quantify the efficiency of a PL FSO receiver. The method is particularly suited to compare PL to other spatial demultiplexers, as it take as a reference a generic compact mode basis for FSO propagation. The method proposed can also be used to quantify the spatial demultiplexer receiver efficiency for more complex use case of FSO communication. We have shown that the modal coupling efficiency $\eta_{m,n}$ can be estimated easily with a near field/far field imaging and with a Gerber-Saxton algorithm. Furthermore, we have also reported an experimental characterization of a PL receiver with a FSO wavefront emulator to validate the numerical approach. The modal decomposition approach predicts faithfully the coupling losses of the PL for a GEO to OGS FSO communication use case. The PL designed shows a -3.3 dB coupling loss in average with limited fading. However, the PL receiver performance are still limited by the number of modes efficiently coupled in the PL. A 19 port PL is currently designed to improve the coupling efficiency of higher order HG modes. We expect such receiver to show even higher coupling efficiency with lower fadings for FSO communications. Implementation of high number of ports PL, associated with photonic integrated beam combination, would paves the way to the next generation of very compact high-speed FSO communication receivers for integrated turbulence mitigation.

**Disclosures**.
The authors declare no conflicts of interest.

**Data availability.**
Data underlying the results presented in this paper are not publicly available at this time but may be obtained from the authors upon reasonable request.